\documentclass[12pt,a4paper]{article}
\usepackage{amsmath}
\usepackage[dvips]{graphicx}
\begin{document}
\begin{titlepage}
\title{Slow down of the mean multiplicity growth at the LHC}
\author{S.M. Troshin, N.E. Tyurin\\[1ex]
\small  \it Institute for High Energy Physics,\\
\small  \it Protvino, Moscow Region, 142281, Russia}
\normalsize
\date{}
\maketitle

\begin{abstract}
We discuss the possibility of changing the energy dependence of the mean multiplicity, i.e. slow down of its   growth  at the LHC energies 
due to a gradual transition to the reflecting scattering mode.
%We  briefly discuss  the possible decoherence observed at the LHC emphasizing that
%the value of the average transverse momentum at  $\sqrt{s}=7$~TeV can be interpreted as a manifestation of the deviation
%from the mechanism of the transient state of rotating matter. 
\end{abstract}
\end{titlepage}
\setcounter{page}{2}

The possibility that the elastic scattering amplitude can exceed the  limitation assumed by the black disk model at very high energies 
was discussed a long time ago \cite{bbdl}.
Such an energy dependence of the amplitude is a manifestation 
of a gradual transition to the reflective scattering mode \cite{intje}. 
The appearance of this mode follows, in its turn, from the fact that opening of the
new inelastic channels with an energy increase, would not lead
to saturation of the total probability of the inelastic collisions at small transverse distance, but instead, it would result in the self-damping
of the inelastic channels \cite{intje,bakbl}.  
The natural question  is how the LHC data
can be interpreted in that sense, namely, are there direct or indirect indications on the presence of this reflective scattering mode at the LHC energies. 
To answer this
question the experimental signatures of the reflective scattering mode  should be considered first.

The straightforward reconstruction of the impact-parameter dependent elastic amplitude is preferable for that purpose. 
It will allow one to conclude on possible crossing of the black disk model limit
for the elastic scattering amplitude.
It requires
careful analysis   of the available experimental data  based on the Fourier-Bessel transformation
and an extra assumption on the real part of the elastic scattering amplitude. 

Another way is to analyse the experimental data on elastic scattering at large transferred momenta (deep-elastic scattering) 
to deduce the possible consequences for the asymptotics of the elastic  amplitude.
Since the deep-elastic scattering probes  the region of small impact parameters,
 it was proposed \cite{deepel}  to use these data 
  for discrimination
of the asymptotic modes  in the hadron scattering. 
In this region the reflective and absorptive scattering modes have  the most significant differences at high energies. 
These two modes are associated with the  different
impact parameter profiles of the inelastic overlap function.
As a result, in the reflective scattering mode  associated with the 
unitarity saturation, the elastic scattering amplitude will asymptotically decouple from the particle production \cite{deepel}.  

At finite energies 
 it should be observed that the deep--elastic scattering 
 has decreasing correlations  with   particle production as the collision energy increases. 
On the other hand the saturation of the black disk limit
implies strong correlation of deep--elastic scattering with the particle production processes.
Respective  asymptotic differential cross--section $d\sigma/dt$
is expected to be four times lower than it is in the case of the reflective scattering mechanism domination.

In this note we consider the implications of the above mentioned decoupling for the global observable related to the many--particle 
production dynamics, namely, the mean multiplicity of the secondary particles $\langle n\rangle (s)$ which is a most general and
transparent quantity related to the particle production processes.

Let us consider the inelastic overlap function 
\[
h_{inel}(s,b)\equiv \frac{1}{4\pi}\frac{d\sigma_{inel}}{db^2} 
\]
 which enters the unitarity equation for the elastic scattering
amplitude $f(s,b)$.  In the impact parameter representation this equation takes simple form
\[
 \mbox{Im} f(s,b)=h_{el}(s,b)+h_{inel}(s,b).
\]
The function $S(s,b)=1+2if(s,b)$ is the $2\to 2$ 
elastic scattering matrix element. 
For simplicity, we consider  the  scattering amplitude $f(s,b)$ to be a pure imaginary function, i.e. 
$f\to if$. The function $S(s,b)$  is a real one in this case, but it can change sign and take  negative values. 
The maximum value of $h_{inel}(s,b)=1/4$
can be reached at high energies at the positive non-zero values of the impact parameter, i.e., for instance, at $b=R(s)$.
The derivatives of $h_{inel}(s,b)$ have the form:
\[
 \frac{\partial h_{inel}(s,b)}{\partial s}=S(s,b)\frac{\partial f(s,b)}{\partial s},\,\,\, 
 \frac{\partial h_{inel}(s,b)}{\partial b}=S(s,b)\frac{\partial f(s,b)}{\partial b}.
\]
It is evident that 
\[
\frac{\partial h_{inel}(s,b)}{\partial b}=0
 \]
at $b=R(s)$,  if $S(s,b)=0$ at this value of the impact parameter. 
Evidently, the derivative of the inelastic overlap function has  the sign opposite to the sign 
of $\partial f(s,b)/\partial b$ in the region where $S(s,b)<0$. It is  the region of the $s$ and $b$ variables
where the function $S(s,b)$ is negative
(the  phase of $S(s,b)$ is such that $\cos 2\delta(s,b)=-1$) is
responsible for the transformation of the central impact--parameter profile of the function $f(s,b)$ into a peripheral profile 
of the inelastic overlap function $h_{inel}(s,b)$ (Fig.1).  
\begin{figure}[h]
%\vspace{}
\begin{center}
\resizebox{8cm}{!}{\includegraphics*{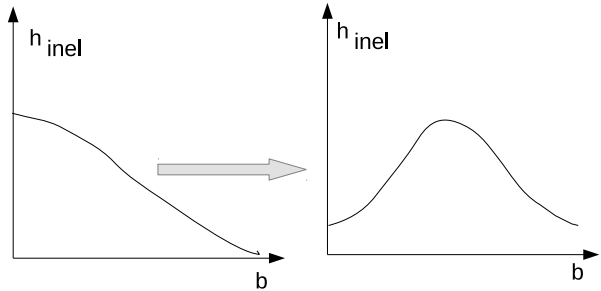}}
\end{center}
\caption[ch1]{Energy evolution of $h_{inel}(s,b)$ from a central to  peripheral profile.}
\end{figure}
It can be easily seen by expressing the function $h_{inel}(s,b)$ as a product, i.e
\[
h_{inel}(s,b)=f(s,b)(1-f(s,b)).
\]
If $f(s,b)>1/2$ at high energy and small impact
parameters, then the function $h_{inel}(s,b)$ will have maximum value 1/4 at the non-zero impact parameter value.
The impact parameter dependence of the function  $h_{inel}(s,b)$ would evolve then with energy from a central to a peripheral one. 
The quantity 
$\langle n\rangle (s)$ is obtained by the integration of the corresponding impact-parameter dependent function 
with the weight function  $h_{inel}(s,b)$, e.g.
the mean multiplicity $\langle n\rangle (s)$ is written in the form
\begin{equation}\label{mm}
 \langle n\rangle (s)=\frac{\int_0^\infty bdb  \langle n\rangle (s,b) h_{inel}(s,b)}
{\int_0^\infty bdb h_{inel}(s,b)}
\end{equation}
The mean multiplicity $\langle n\rangle (s,b)$ in the geometric approach has a central dependence on $b$ (cf. e.g. \cite{chya}).
It is often considered as a folding integral 
\begin{equation}\label{ns}
 \langle n\rangle (s,b)=n_{0}(s)D_{1}\otimes D_{2},
\end{equation}
where $D_i$ are the two-dimensional impact-parameter dependent matter distributions  in the colliding hadrons. 
Therefore, when the
 weight function $h_{inel}(s,b)$ evolves with energy to a peripheral profile, the energy dependence of  $\langle n\rangle (s)$ should start
 to slow down. For example, it can be expected that power-like energy dependence would gradually slow down, namely  
 \[
 s^\beta \to
 s^{\beta-\Delta}
 \]
 starting in  the energy region where
 the function $S(s,b)$  takes negative values. The parameters $\beta$ and $\Delta$ ($\beta > \Delta$) are determined by the parameters 
 of the model, e.g.  \cite{mult}.
Proceeding from the available experimental information, one can roughly estimate the starting energy of slow down 
of  $\langle n\rangle (s)$  in the range  of  $\sqrt{s}=3-5$ TeV. 
 This estimate is based on the fact  
that the value of $\mbox{Im} f(s,b=0)$ increases
from $0.36$  (CERN ISR) to $0.492\pm 0.008$ (Tevatron)  and has a value which is very close to the black disk limitation $0.5$ at 
$\sqrt{s}=2$ TeV \cite{girom}. 
If it is so, it would testify  in favor of a gradual transition to the reflective elastic scattering starting already at the LHC energies. 

Unfortunately, the data for the mean
multiplicity at the LHC are available for the central region of the rapidity only \cite{cms,atlas,alice}. 
Straightforward extrapolation of this energy dependence to the whole
region of rapidity at any fixed energy \cite{block}  looks oversimplified.  The data in the regions not covered by the  
measurements  could demonstrate different energy dependences. However, assuming  the above extrapolation valid, one should 
expect the slowing mean multiplicity growth be shifted to the region of $\sqrt{s}=10-15$ TeV.  This conclusion is valid also in the 
particular model based on assumption of the Eq. (\ref{ns}) for the mean multiplicity distribution $\langle n\rangle (s,b)$. For the 
function $\langle n\rangle (s)$ the following relation is valid
\[
 \langle n\rangle (s,b)=n_{0}(s)F(s),
\]
where the function $F(s)$ can be calculated in a  similar way to the calculation of the gap survival probability in the double-pomeron exchange processes 
performed in \cite{dpe}.
The  factor $F(s)$ has a maximum at $\sqrt{s}=10-15$ TeV and  decreases beyond those energies like a negative power of energy.

Thus, one can state that the gradual transition to the
 reflective scattering mode (with a prominent peripheral form of the overlap function  $h_{inel}(s,b)$) would lead
 to suppression of the higher multiplicity events in the  region of the small transverse distances. 
 Asymptotically, only reflective elastic scattering would survive
 at small values of the impact parameter.
 A standard assumption in many geometrical approaches is that  the distribution of the 
 mean number  of the secondary particles over impact parameter 
 is  supposed to have a maximum in the region $b\simeq 0$.  
 Combination of these two facts is translated (due to integration over impact
 parameter) to  slowing down energy dependence of the mean multiplicity $\langle n\rangle (s)$  at the energies
 when the inelastic overlap function $h_{inel}(s,b)$ starts to be peripheral. 
 The experimental measurements of the mean multiplicity in the energy region of $\sqrt{s}=10-15$ TeV would be interesting and  helpful for
 discrimination of the different modes of hadron interaction and would provide hints for the asymptotics.

\section*{Acknowledgement}
We are grateful to Sergey Sadovsky  for 
discussion on the experimental data for the mean multiplicity obtained by the ALICE Collaboration at the LHC.

\end{document}